\documentclass{iopconfser}

\newcommand{\be}{\begin{equation}}

\newcommand{\ee}{\end{equation}}

\newcommand{\bea}{\begin{eqnarray}}

\newcommand{\eea}{\end{eqnarray}}

\usepackage{a4wide,amssymb,cite,graphicx}
\usepackage{epsfig}
\usepackage{hyperref}
\usepackage{tabularx}
\usepackage{epsfig}

\usepackage{subfigure}

\begin{document}

\noindent Conference Proceedings for BCVSPIN 2024: Particle Physics and Cosmology in the Himalayas\\Kathmandu, Nepal, December 9-13, 2024 

\title{Exploring generalized Starobinsky
Model of Inflation: Observational Constraints}

\author{Saisandri Saini$^{1}$, Akhilesh Nautiyal$^{2}$}

\affil{$^1$Department of Physics, Malaviya National Insitute of Technology, Jaipur, India}
\affil{$^2$Department of Physics, Malaviya National Insitute of Technology, Jaipur, India}

\email{2019rpy9082@mnit.ac.in}
\email{akhilesh.phy@mnit.ac.in}

\begin{abstract}
We examine the power-law Starobinsky model, a generalized version of the Starobinsky inflation model, characterized by a power-law correction to Einstein gravity. Employing the $f(R)$ formalism, the scalar and tensor power spectra were numerically computed as functions of the dimensionless parameters $M$ and $\beta$. A Markov Chain Monte Carlo (MCMC) analysis was conducted using Planck-2018, BICEP3 and BAO observational data, yielding precise constraints on 
$\beta = 1.987^{+0.013}_{-0.016},\, \, \, 95\%\, C.\, L.$. and $ \log_{10}M = -4.72^{+0.21}_{-0.20}$. The derived scalar spectral index $n_s=0.9676^{+0.0069}_{-0.0068}$ and tensor-to-scalar ratio $r=0.0074^{+0.0061}_{-0.0044}$ lie within the bounds set by Planck observations. We analyze a general reheating scenario while keeping the number of e-folds during inflation, $N_{pivot}$, fixed. The analysis confirms that deviations from the Starobinsky $R^2$ model are observationally viable, with implications for high-energy physics and supergravity-based inflationary models.
\end{abstract}

\section{Introduction}
The inflationary paradigm \cite{Guth:1980zm} provides a robust framework for addressing the shortcomings of the standard Big Bang cosmology, such as the horizon and flatness problems \cite{Linde:1981mu}, while simultaneously offering a mechanism for the origin of cosmic microwave background (CMB) anisotropies and large-scale structure \cite{Mukhanov:1981xt,Starobinsky:1982ee,Guth:1985ya}. Among various inflationary scenarios\cite{Martin:2013tda}, the Starobinsky 
$R^2$ model \cite{Starobinsky:1980te} stands out as one of the most compelling in analysis of recent Planck 2018 results \cite{Planck:2018jri} . It relies solely on a curvature-squared term within the $f(R)$ framework, without requiring the introduction of additional scalar fields.

Despite its success, generalizations of the Starobinsky model have gained attention, particularly those incorporating power-law corrections to the curvature term \cite{Muller:1989rp,Gottlober:1992rg}. Such extensions not only allow deviations from the canonical $R^2$ form but also provide a fertile ground for connecting inflationary physics with high-energy frameworks like supergravity \cite{Ellis:2013nxa,Ellis:2018zya}.

In this work, we focus on the power-law Starobinsky model, where the action includes a term proportional to $R^{\beta}$. This generalization introduces two free parameters, $M$ and $\beta$, which are subject to observational constraints. By numerically evaluating the scalar and tensor perturbations, we perform a Bayesian parameter estimation using the latest CMB and BAO data to assess the observational viability of this model. The derived constraints on $\beta$ offer new insights into the parameter space beyond the Starobinsky $R^2$ limit, highlighting the potential for deviations and their implications for inflationary dynamics and reheating scenarios. The analyses in  \cite{Chakravarty:2014yda,Motohashi:2014tra,Odintsov:2022bpg,Meza:2021xuq} using the slow-roll approximation also found a deviation from $\beta=2$. In \cite{Saini:2023uqc}, derived constraints on the inflaton potential parameters $M$ and $\beta$ from equation \ref{pot} using Planck-2018 data, in combination with BICEP3 \cite{BICEP:2021xfz}, (BAO), (DES) and Pantheon observations. A general reheating scenario is considered, along with variations in the number of e-folds during inflation, $N_{pivot}$. However, in this study, we have set $N_{pivot}=50$ as a fixed value.

\section{The Generalized Starobinsky Model of Inflation}
\subsection{Overview of the Model}
The generalized Starobinsky model introduces a power-law correction to the 
$f(R)$ framework of gravity, extending the well-known 
$R^2$ inflationary model. This approach explores deviations from the canonical Starobinsky potential, offering a broader parameter space for analyzing inflationary dynamics and connecting with high-energy physics scenarios. The action for the model in the Jordan frame is given by \cite{Chakravarty:2014yda} :
\begin{equation}
S_J = \frac{-M_{Pl}^2}{2}\int \sqrt{-g} \left(R+\frac{1}{6M^2}\frac{R^{\beta}}{M_{Pl}^{2\beta-2}}\right) d^4 x. \label{SJ} 
\end{equation}
where $M_{Pl}^2 = (8\pi G)^{-1}$, g is the determinant of the metric $g_{\mu \nu}$, R is the Ricci Scalar and $M$ is a dimensionless real 
parameter.

For $\beta=2$, the model reduces to the original Starobinsky $R^2$ inflationary model.

\subsection{Transformation to the Einstein Frame}
The formulation in the Jordan frame can be re-expressed in the Einstein frame by a conformal transformation. This allows the action to take the standard Einstein-Hilbert form with an additional scalar field $\chi$. The transformed action in the Einstein frame is:
\begin{equation}
S_E = \int d^4x\sqrt{-\tilde{g}}\left(\frac{-M_{Pl}^2}{2}\tilde{R}+\frac{1}{2}\tilde{g}^{\mu \nu}\partial_{\mu}{\chi}\partial_{\nu}{\chi}+U(\chi)\right),\label{SE}
\end{equation}
Where $\tilde{g}$ and $\tilde{R}$ are the metric and Ricci scalar in the Einstein frame. $U(\chi)$ is the scalar potential derived from the original $f(R)$ function.

The scalar field potential in the Einstein frame for the generalized Starobinsky model is given as:
\begin{equation}
U({\chi}) = (\frac{{\beta}-1}{2})\left(\frac{6M^2}{{\beta}^{\beta}}\right)^{\frac{1}{\beta-1}} \exp\left[{\frac{2\chi}{\sqrt 6}\left(\frac{2-\beta}{\beta-1}\right)}\right]\times \left(1-\exp\left(\frac{-2\chi}{\sqrt 6}\right)\right)^\frac{\beta}{\beta-1}. \\ \label{pot}
\end{equation}

\subsection{Background and Perturbation Equations}
In the Einstein frame, the equations governing the background dynamics during inflation are:
\textbf{Friedmann equations:}
\bea
H^2 &=& \frac{1}{3M_{Pl}^2}\left[\frac{1}{2}\dot{\chi^2} + V(\chi)\right]. \label{H2}\\
\dot H &=& -\frac{1}{2 M_{Pl}^2} \label{hprime}
\eea
\textbf{Scalar Field Equation:} 
\be
\ddot{\chi} + 3H\dot{\chi} + \frac{dV(\chi)}{d\chi} = 0. \label{evo}
\ee
Here, the dot represents the differentiation with respect to cosmic time. 

During inflation, perturbations in the metric and scalar field give rise to observable quantities such as the scalar power spectrum $P_s$ and and the tensor-to-scalar ratio $r$. These perturbations are described by the Mukhanov-Sasaki equation for scalar perturbations and the tensor perturbation equation \cite{Mukhanov:1988jd,Sasaki:1986hm} . \\
\textbf{Mukhanov-Sasaki Equation (Scalar Perturbations):} 
\be
\frac{d^2 u_k}{d\tau^2} + \left(k^2 - \frac{1}{z}\frac{d^2z}{d\tau^2}\right)u_k = 0,\label{uk}
\ee
where $z=\frac{1}{H}\frac{d\chi}{d\tau}$ encodes the background dynamics, $\tau$ denotes the conformal time. $k$ is the comoving wavenumber.
The scalar power spectrum is expressed as:
\be
\mathcal{P_\mathcal{R}}(k) = \frac{k^3}{2\pi^2}\bigg|\frac{u_k}{z}\bigg|^2, \label{Ps}
\ee
\textbf{Tensor Perturbations:}
Similarly the mode equation for tensor perturbations generated during inflation is given as
\be
\frac{ dv_k}{d\tau^2} + \left(k^2 - \frac{1}{a}\frac{d a^{\prime\prime}}{d\tau^2}\right)v_k = 0,\label{vk}
\ee
and the primordial tensor power spectrum is given as
\be
\mathcal{P}_{t}(k) = \frac{4}{\pi^2}\frac{k^3}{M_{Pl}^2}\bigg|\frac{v_k}{a}\bigg|^2.\label{Pt}
\ee
The scalar spectral index $n_s$ and the tensor spectral index $n_t$ are  computed numerically from the power spectra by applying their respective definitions \cite{Bassett:2005xm}.
\bea
n_s &=& 1 + \frac{d \ln\mathcal{P_\mathcal{R}}}{d \ln k},\label{ns}\\
n_t &=& \frac{d \ln\mathcal{P}_{t}}{d \ln k},\label{nt}
\eea
The tensor-to-scalar ratio $r$ is defined by \cite{Bassett:2005xm}
\be
r = \frac{\mathcal{P}_{t}}{\mathcal{P_\mathcal{R}}}.\label{r}
\ee

The background and perturbation equations are solved numerically using the e-folding number ,  $N = \ln a $ as the
independent variable. The initial conditions are set deep in the slow-roll regime, where the scalar field's velocity is small, and the perturbation modes are initialized in the Bunch-Davies vacuum state.
The Planck CMB observations impose constrains on $n_s$ and $r$. However, in this study  $n_s$ and $r$ are treated as derived quantities, while the primary parameters constrained directly by the CMB data are those of the inflaton potential (\ref{pot}), namely $M$ and $\beta$.
\section{Observational Results and Parameter Constraints}
In this study, the parameters of the power-law Starobinsky model were constrained using data from Planck-2018, BICEP3 and baryon acoustic oscillations (BAO). The analysis was conducted using a numerical framework to solve the background and perturbation equations of the model without relying on the slow-roll approximation. To explore the parameter space of the model, the publicly available ModeCode \cite{Mortonson:2010er} was adapted to incorporate the power-law Starobinsky potential. This numerical tool computed the scalar and tensor power spectra, which were then input into CAMB \cite{Lewis:1999bs} to generate the angular power spectra for the cosmic microwave background (CMB) anisotropy and polarization. CosmoMC \cite{Lewis:2002ah} was employed for the Markov Chain Monte Carlo (MCMC) analysis, enabling robust parameter estimation.

\begin{table}[h]
\centering
\begin{tabular}{| c| c |c| c |}
\hline
 Parameter &  68\% limits & 95\% limits   &  99\% limits  \\
\hline

{\boldmath$N_{pivot}$} & $48.7^{+2.9}_{-7.8}  $ & $49^{+10}_{-8}            $ & $49^{+10}_{-8}             $\\
\hline

{\boldmath$\log_{10} M$}  & $-4.72^{+0.11}_{-0.14}  $ & $-4.72^{+0.21}_{-0.20}     $ & $-4.72^{+0.27}_{-0.23}     $\\
\hline

{\boldmath$ \beta$} & $1.987^{+0.011}_{-0.0048}  $ & $1.987^{+0.013}_{-0.016}   $ & $1.987^{+0.014}_{-0.022}   $\\
\hline

{\boldmath$n_s$} & $0.9676\pm 0.0035          $ & $0.9676^{+0.0069}_{-0.0068}$ & $0.9676^{+0.0090}_{-0.0089}$\\
\hline
 
{\boldmath$r$}  & $0.0074^{+0.0016}_{-0.0039}$ & $0.0074^{+0.0061}_{-0.0044}$ & $0.0074^{+0.0086}_{-0.0048}$\\
\hline
\end{tabular}
\caption{Constraints on parameters of potential, $r$ and $n_{s}$ using Planck-2018,
BICEP/Keck (BK15) and BAO observations.}
\label{Tab:constraint}
\end{table}

Table \ref{Tab:constraint}  presents the constraints derived for the parameters of potential \ref{pot}, including the e-foldings $N_{pivot}$ and the derived parameters $r$ and $n_s$. The priors for the model parameters were set same as \cite{Saini:2023uqc} as follows:

\begin{equation}
    \log_{10} M \in [-6.5, -3.0], \quad \beta \in [1.90, 2.07].
\end{equation}

The Table \ref{Tab:constraint} clearly indicates that the most suitable value of $\beta$ is
\be
\beta = 1.987^{+0.013}_{-0.016},\, \, \, 95\%\, C.\, L., \label{betadata}
\ee

\begin{figure}[htbp]
\begin{center}
\subfigure[]{
 \includegraphics[width=7cm, height = 7cm]{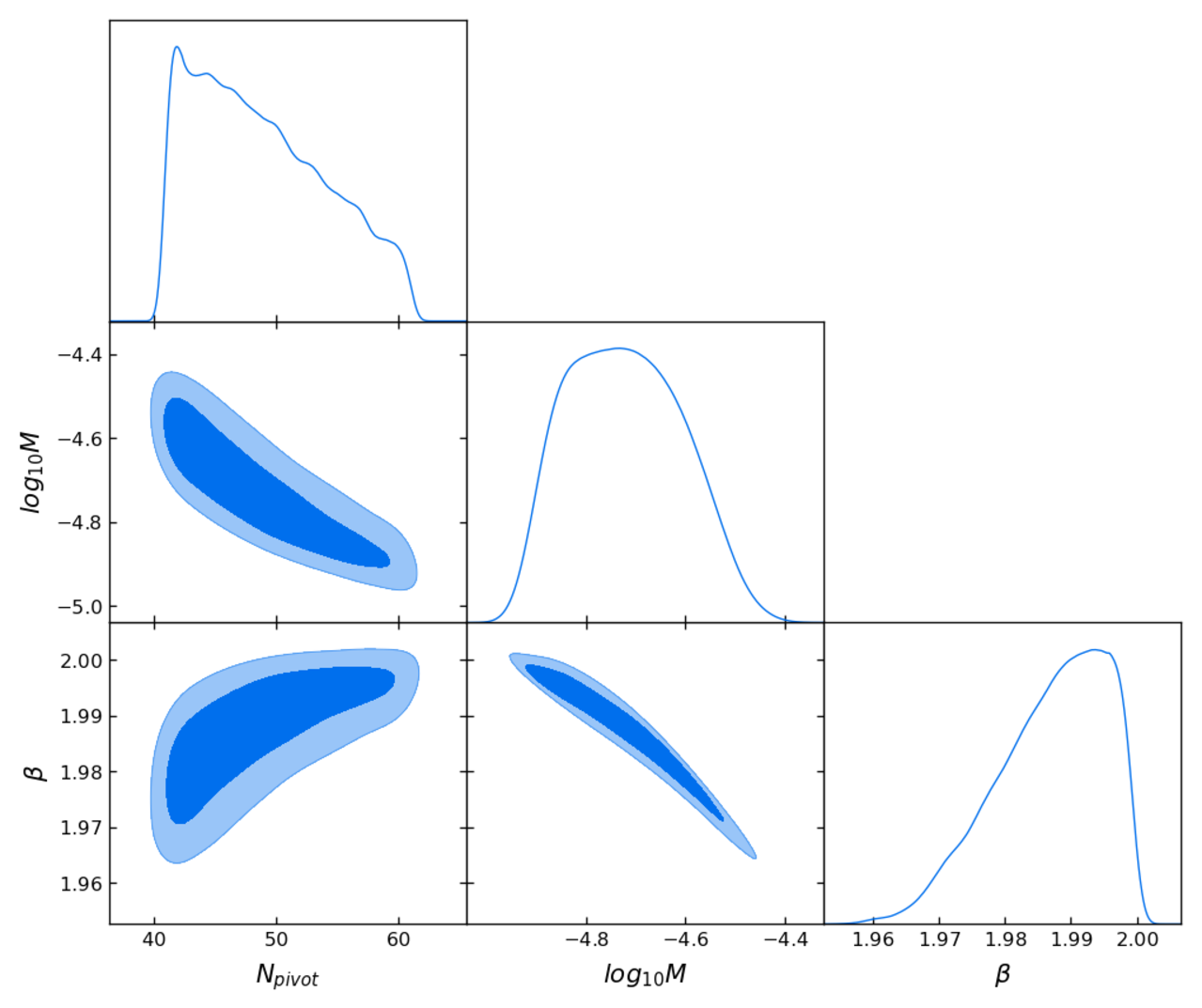}
 \label{fig:plot}
}
\subfigure[ ]{
\includegraphics[width=7cm, height = 7cm]{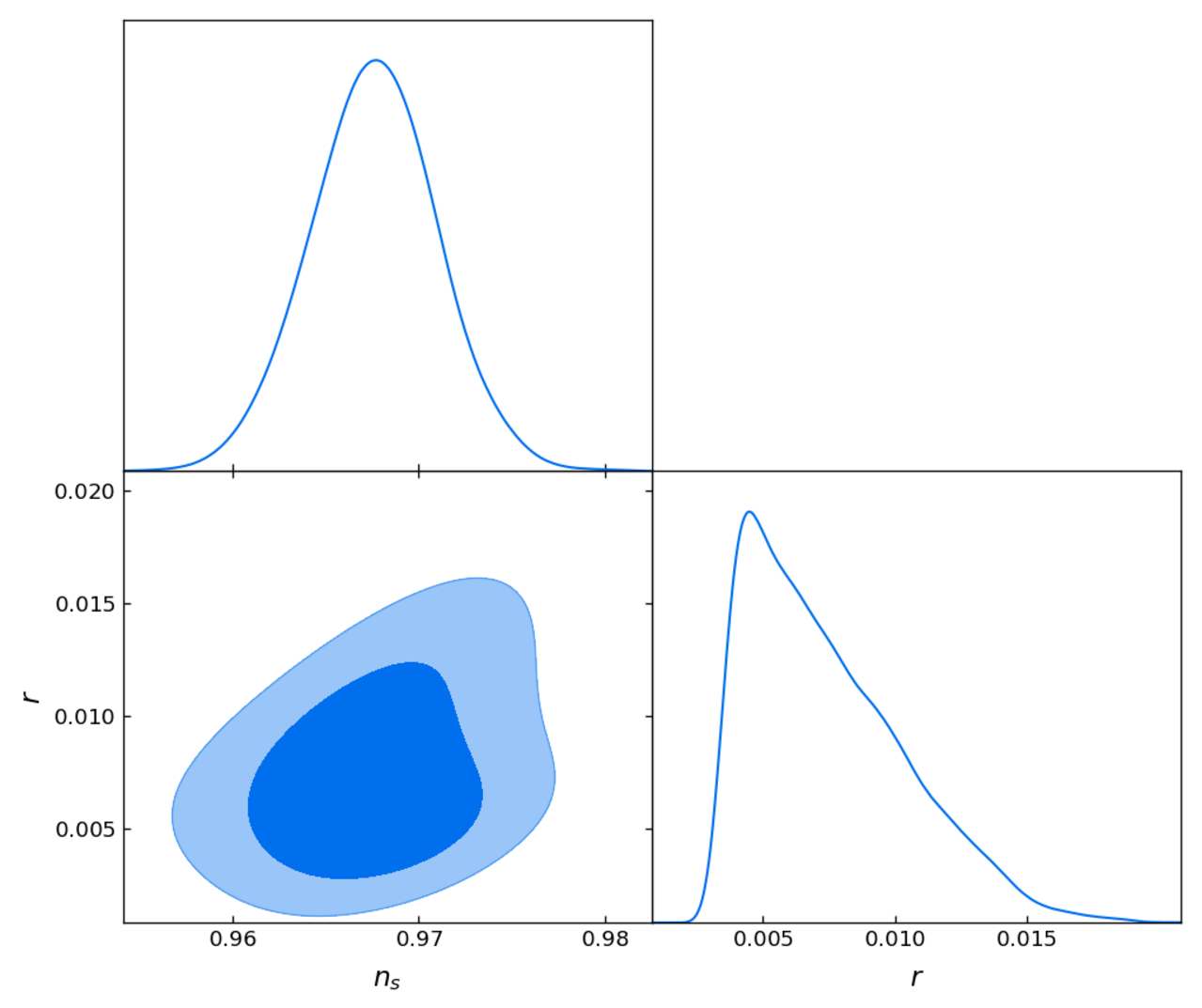}

 \label{fig:nsr}
}
\caption{Marginalized joint two-dimensional $68\%$ C.L. and $95\%$ C.L.  
constraints on parameters of potential (a) and $N_{pivot}$ (b) using Planck-2018, 
BICEP/Keck (BK15) and BAO data}
\label{fig:triangularplot}
\end{center}
\end{figure}
The triangular plots \ref{fig:triangularplot} display the posterior distributions and correlation contours between different cosmological parameters. Fig. \ref{fig:plot} indicates a strong correlation between model parameters $(\log_{10} M, \beta)$ and $N_{pivot}$ also to each other. Fig.\ref{fig:nsr}  the spectral index $n_s$ and the tensor-to-scalar ratio $r$ display a strong correlation. As observed in Table \ref{Tab:constraint},  the best-fit values of the scalar spectral index $n_s$ and the tensor-to-scalar ratio $r$, derived from the optimal potential parameters of the power-law Starobinsky model, remain consistent with the Planck limits.

In this study, we investigate power law Starobinsky inflation in the context of Planck-2018, BICEP3 \cite{BICEP:2021xfz} CMB data. We consider the inflaton potential (\ref{pot}) for power-law Starobinsky inflation in the Einstein frame and numerically compute the scalar and tensor power spectra using MODECODE. Utilizing these results, we conduct an MCMC analysis with COSMOMC to constrain the inflaton potential parameters $\beta$ and $M$, as well as the number of e-folds at the pivot scale, $N_{pivot}$. To account for deviations from Starobinsky inflation, we vary $\beta$ within the range $1.9$ to $2.07$. Based on Planck-2018 data, we find that 
$\beta = 1.987^{+0.013}_{-0.016},\, \, \, 95\%\ $ confidence level ($\beta = 1.966^{+0.035}_{-0.042},\,$ in \cite{Saini:2023uqc}), indicating that current CMB and LSS observations favor a slight deviation from the standard Starobinsky inflationary scenario.

\bibliographystyle{jhep}
\bibliography{Replace with name of your bib file}{}

\end{document}